\documentclass[aps,prb,twocolumn,showpacs]{revtex4}
\usepackage{graphicx,epstopdf}
\usepackage{amssymb,amsmath}
\usepackage{simplewick}
\usepackage{epsf,epsfig}
\newcommand{\dg}{{^{\dagger }}}

\newcommand{\pmat}[1]{\begin{pmatrix} #1 \end{pmatrix}}
\newlength{\bxwidth}\bxwidth=2.0 truein
\newcommand\frm[1]{\epsfig{file=#1,width=\bxwidth}
}
\begin{document}
\title{End states in a 1-D topological Kondo insulator}
\author{Victor Alexandrov$^1$ and Piers Coleman$^{1,2}$
\\
$^{1}$Center for Materials Theory, Department of Physics and Astronomy,
\\
Rutgers University, Piscataway, NJ 08854-8019, USA
\\
$^{2}$ Department of Physics, Royal Holloway, University
of London,
\\ Egham, Surrey TW20 0EX, UK.
}

\begin{abstract}
To gain further insight into the properties of interacting topological 
insulators, we study a 1 dimensional model of topological Kondo
insulators which can be regarded as the strongly interacting 
limit of the Tamm-Shockley model.
Treating the model in a large $N$ expansion, 
we find a number of competing ground-state solutions, including
topological insulating and valence bond ground-states.  One of the
effects to emerge in our treatment is a 
reconstruction of the Kondo screening process near the boundary of the
material (``Kondo band bending'' ). 
Near the boundary for localization into a valence bond state, we find
that the conduction character of the edge state grows substantially, 
leading to states that extend deeply into bulk. We speculate that such
states are the one-dimensional analog of the light f-electron surface
states which appear to develop in the putative topological Kondo insulator, 
SmB$_{6}$. 
\end{abstract}
\maketitle
\section{Introduction}

Topological insulators \cite{Haldane1988,Fu2005,Kane2005,Kane2005_2,Fu2007, Moore2010,TIreview1,TIreview2,Moore2007,Roy2009,Qi2008} have attracted
great  attention as a new class of band insulator
with gapless surface or edge states, robustly protected by combination
of time-reversal symmetry and the non-trivial topological winding of
the occupied one-particle wavefunctions. 
The surface states of a topological insulator are ``massless''
excitations carried by an 
odd number of Dirac cones in the Brillouin zone.

Various proposals have been made for 
strongly correlated electron analogues\cite{Wang2010,Dzero2010,Jiang2012,Nourafkan2013,Wang2012,Fidkowski2011,Wang2014} of topological band
insulators. To date the best candidate strongly correlated topological
insulator is SmB$_{6}$, a local moment metal which transforms into a 
Kondo insulator, once the moments screen  at low temperatures
 ($<70K$) \cite{Geballe1969}. This material was first predicted to
be a topological Kondo insulator \cite{Dzero2010}
 % Many aspect of its properties are topological. 
and recently shown to exhibit conducting in-gap surface states,
which develop below $4$K \cite{exp1smb6,exp2smb6,exp3smb6}. 
While these results 
are consistent with a topological Kondo insulator, 
a definitive observation of Dirac cone excitations with polarized
quasiparticles has not yet been reported. However, tentative data of
the Dirac cone surface states have become available in both Quantum
oscillation \cite{QuantumOsc2013} and ARPES measurements\cite{Hasan2013,ARPES2,ARPES3,ARPES4}. One of the unexpected
features of these measurements is the presence of ``light'',
high-velocity surface quasiparticles, with the Dirac point far outside
the gap. These tentative results are puzzling, because their group
velocities appear 10 to 100 times larger than that expected
in a heavy fermion band
\cite{Dai2013,Alexandrov2013}.

\begin{figure}[h!]  \centering
\includegraphics[width=0.9\linewidth]{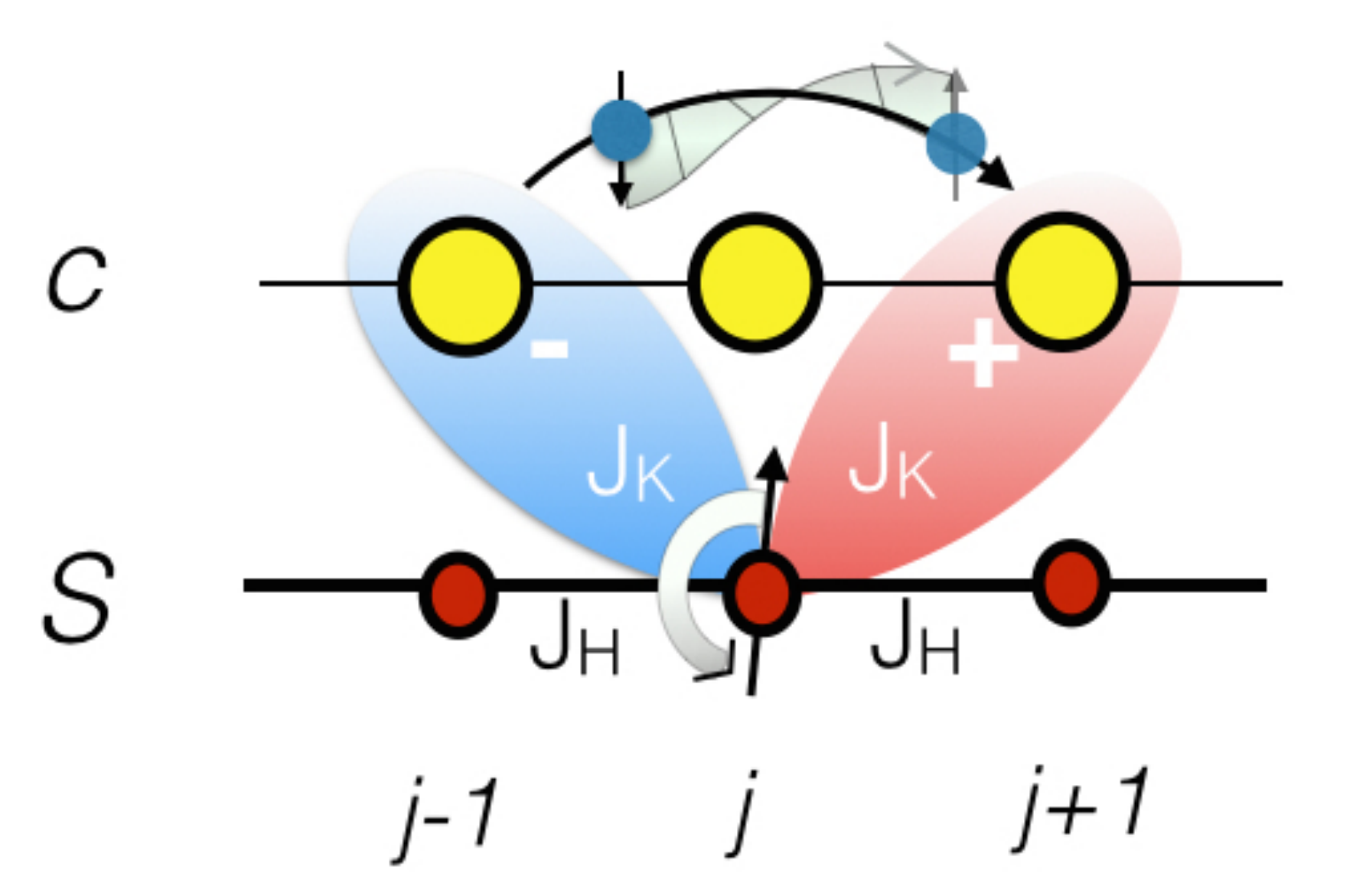}
\caption{(Color online) Schematic illustration of the 1D p-wave Kondo
insulator Hamiltonian (\ref{theModel}). The model contains a chain of Heisenberg
spins coupled by an antiferromagnetic nearest neighbor
Heisenberg coupling, plus a tight-binding chain of
conduction electrons (c).  Each localized moment 
is coupled to the conduction sea via a Kondo ``cotunneling'' term that
exchanges spin between a localized moment at site $j$ 
and a p-wave combination of conduction electron states formed between neighboring sites $j-1$ and $j+1$.
}.  \label{fig:fig_model}
\end{figure}

These results provide motivation for the current paper.  Here we
introduce a simple one dimensional ``p-wave Kondo lattice'' which
gives rise to a topological Kondo insulator 
that can be studied by a variety of methods. In this initial study we
carry out the simplest mean-field treatment of our model, an approach
which is technically exact in the large $N$ limit, using it to
gain insight into the nature of the edge states and to propose
variational ground-states for the model. This work is also an
important warm-up exercise for a three dimensional
model. 
The model is schematically depicted in  Fig. \ref{fig:fig_model}. It can be regarded  as a strongly interacting limit of Tamm-Shockley model \cite{Tamm1932,Shockley1939,Yakovenko2012}. 

A second goal of this work is to gain insight into the 
the impact of the boundary on the Kondo effect, a phenomenon we refer
to as ``Kondo band bending''. In the conventional Kondo insulator
model, the
hybridization between local moments and conduction electrons is local
and the strong-coupling ground state involves a 
Kondo singlet at every site, with
minimal boundary effects.  In the case of non-trivial
topology the Kondo singlets are non-local objects
(Fig. \ref{fig:fig_model}) which are partially broken at the
boundary. We seek to understand how this influences the Kondo effect
and the character of the edge states at the boundary.

\section{The model}

Our model describes conduction electron fluid interacting with
an antiferromagnetic Heisenberg spin chain 
via a Kondo co-tunneling term with p-wave character. The Hamiltonian is given by 
\begin{equation}\label{theModel}
H = H_c +H_H + H_K,
\end{equation}
where
\begin{align}
H_c&= -t\sum_{j, \sigma }    (c_{j+1\sigma }^\dag c_{j\sigma } + \rm H.c)
\\
\label{H_H} 
 H_H&= J_H \sum_i {\bf S}_j \cdot {\bf S}_{j+1},
\\
 \label{H_K}
H_K &= \sum_{j,\alpha\beta} \frac{J_K{(j)}}{2} {\bf S}_j\!\cdot\!
p_{j,\alpha
}^\dag {\boldsymbol\sigma}_{\alpha \beta }
p_{j,\beta}.
\end{align}
Here the site index runs over the length of the chain, $j \in [1,L]$,
$t$ is the nearest neighbor a hopping matrix element, $J_H$ is a
nearest neighbor Heisenberg coupling and $J_K (j)$ is the Kondo
coupling at site $j$.  The chemical potential of the conduction
electrons has been set to zero, corresponding to a half-filled
conduction band. In contrast to the conventional 's-wave' Kondo model,
the Kondo effect is non-local.  In particular, the electron Wannier
states that couple to the local moment have p-wave symmetry
\begin{equation}
p_{j,\sigma} \equiv c_{j+1,\sigma}-c_{j-1\sigma}.
\end{equation}
The Kondo coupling now permits the process of 
``co-tunneling'' whereby an electron can hop across a spin as it flips
it. 
The odd-parity co-tunneling terms
are  a consequence of the underlying 
hybridization with localized p-wave orbitals.
When this hybridization is
eliminated via a Schrieffer-Wolff transformation, the resulting Kondo
interaction contains an odd-parity form factor.

The boundary spins
have a lower connectivity, giving rise to a lower Kondo temperature
which tends to localize them into a magnetic state. 
To examine these effects in greater detail, 
we take the Kondo coupling $J_{K} (j)=J_{K}$ to be
uniform in the bulk, but to have strength $\alpha J_{K}$ at the 
boundary, 
\begin{equation}
J_K(j)= \left\{\begin{array}{ll}
\alpha J_K & \hbox{endpoints}\ \ ( j = 1 \hbox{\ or\ }  L), \\ 
 J_K  & \hbox{bulk}\ (j\in [2,L-1] ).
\end{array} \right.
\end{equation}
By allowing the
end couplings to be enhanced by a factor $\alpha $ we can crudely
compensate  for the localizing effect of the reduced boundary
connectivity. In real 3D Kondo insulators, this surface enhancement effect
(``Kondo band-bending'') would occur in response to 
changes in the valence of the magnetic ions near the surface. 
For Sm and Yb Kondo insulators, the valence of the surface ions is
expected to shift to a more mixed valent configuration, enhancing
$\alpha $, while in Ce Kondo insulators, the opposite effect is
expected. 

To formulate the model as a canonical field theory, we rewrite the spin ${\bf S}_{j}$
using Abrikosov pseudo-fermions $f_{j\sigma}$, 
as 
\begin{equation}
{\bf S}_{j} = \sum_{\sigma\sigma'} f\dg _{j\sigma}
\boldsymbol{\sigma}_{\sigma\sigma'}f_{j\sigma'}, 
\end{equation}
with the associated ``Gutzwiller'' constraint $n_{f,j}=1$ at each site. 
 After applying the completeness relations for the Pauli matrices 
in (\ref{H_K}) we obtain the Coqblin-Schrieffer form of the Kondo
interaction, 
\begin{equation}
H_K = -J_K\sum_{j,\alpha\beta}\left( f^\dag_{j, \alpha} p_{i\alpha } \right) \left( p^\dag_{j, \beta} f_{i\beta } \right)
\end{equation}
where we have imposed the constraint. 
In an analogous fashion, the local moment interaction (\ref{H_H}) can be
re-written as 
\begin{equation}
H_H= - J_H \sum_{j, \alpha\beta}  (f^\dag_{j+1, \alpha} f_{j\alpha })
(f^\dag_{j, \beta} f_{j+1\beta }) \end{equation}
If we now cast the Hamiltonian inside a path integral, we can
factorize the Kondo and Heisenberg interactions using a 
Hubbard-Stratonovich decoupling, 
\begin{eqnarray}\label{mf_hamiltonian}
&&H\rightarrow H_{c}+ \sum_{j,\sigma}
\left( 
 \bar  V_{j} 
  (c\dg _{j+1\sigma }-c\dg_{j-1\sigma })f_{j \sigma}
  + {\rm H.c} \right) +\frac{|V_{j}|^2}{J_K{(j)}}  
\cr
&&\qquad+ \sum_{j, \sigma }\left( 
 \Delta_{j}  f^\dag_{j+1,\sigma} f_{j, \sigma} +\hbox{H.c}
\right)
 +{|\Delta_{j}|^2\over J_H} \cr
&&\qquad +\sum_j \lambda_{j} (n_{f,j}-1),
\end{eqnarray}
with the understanding that auxiliary fields  $V_i$, $
\Delta_{j}$ and $\lambda_{j}$
are fluctuating variables, integrated within a path
integral.  The last  term imposes the constraint $n_{f,j}=1$ at each
site.

 In this formulation of the problem $V_i$ determines the Kondo
 hybridization on site $i$ and $\Delta_i$ is the order parameter for
 RVB-like state formed on the link $i$ between local moments.
In translating our mean-field results back into the physical subspace
of spins and electrons it is important to realize that 
the f-electron operators (which are absent in the original spin
formulation of the model) represent composite fermions that result
from the binding of spin flips to conduction electrons as part of the
Kondo effect. By comparing
(10) with (4), we see that the f-electron represents the following
(three-body) contraction between conduction and spin operators:
\begin{eqnarray}\label{l}
\contraction{}{\bf S}{_j\cdot{\boldsymbol\sigma}_{\alpha\beta}\hskip -1mm}{p}
{\bf S}_j\cdot
{\boldsymbol\sigma}_{\alpha\beta}p 
_{j,\beta } 
&\equiv &\left(\frac{2\bar V }{J_{K}} \right)f_{j,\alpha },\cr
\contraction[2ex]{}{p\dg }{_{j,\beta } {\boldsymbol{\sigma}}_{\beta\alpha}\cdot\hskip -1mm}{{\bf S}}
p\dg_{j,\beta } {\boldsymbol{\sigma }}_{\beta \alpha }\cdot{\bf S}
_{j}
&\equiv & \left(\frac{2 V }{J_{K}} \right)f\dg _{j,\alpha }.
\end{eqnarray}
At low energies, these bound-state objects behave as independent
electron
states, injected into the conduction sea to form a filled band and
create a Kondo insulator. 
%
%for  the purpose of simplicity here and the remaining   we assume particle-hole symmetry. Due to the symmetry both chemical potentials are zero, $\mu_c = \lambda_j = 0$. 

\subsection{Homogeneous mean field approximation}

\begin{figure}[h!]
\centering
\includegraphics[width=.7\linewidth]{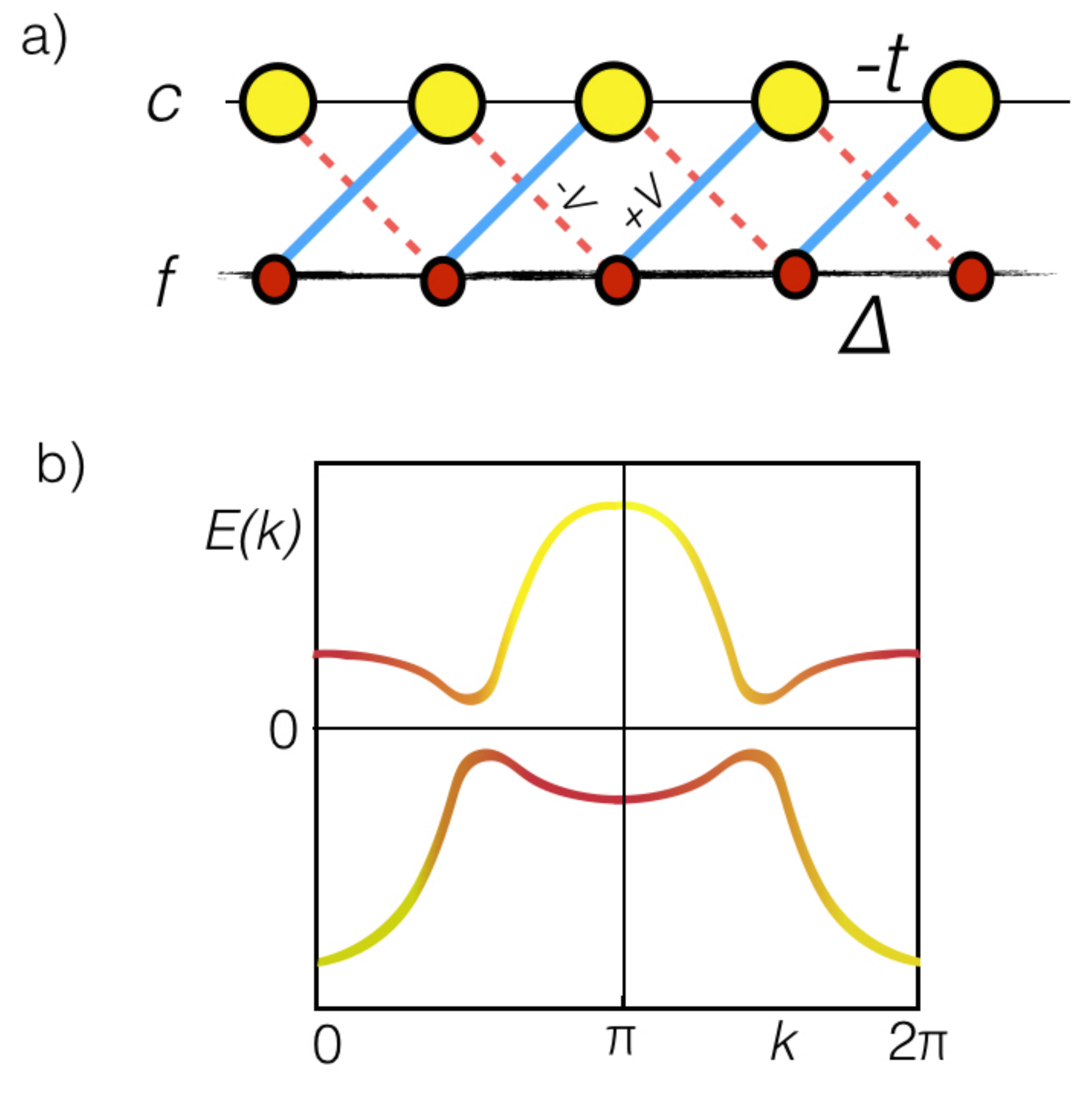}
\caption{(Color online) Illustrating the tight-binding model. (a) Real
space structure. (b) Dispersion of quasiparticles, showing band
inversion at $k=\pi$(\ref{TB model})
}
\label{fig:fig_TB}
\end{figure}

In the homogeneous
 mean field treatment of the Hamiltonian
(\ref{mf_hamiltonian}) we assume that the bulk fields $V_{j}$, $\Delta_{j}$
and $\lambda_{j}$ are constants. The saddle-point Hamiltonian then
becomes a translationally invariant tight binding model. For periodic
boundary conditions, taking $\Delta_{j}=\Delta $, $V_{j}=V$, 
we obtain
\begin{equation}\label{}
H = H_{TB} + L\left(\frac{\vert V\vert^{2}}{J_{K}}+ 
\frac{\vert \Delta \vert^{2}}{J_{H}}- \lambda  \right).
\end{equation}
$H_{TB}$  can be written in momentum space as 
%$H_{TB} = \sum_k H(k)$ where
\begin{equation}\label{TB model}
H_{TB}=\sum_{k}(c_{k\sigma}^\dagger, f_{k\sigma}^\dag) 
\overbrace {\begin{pmatrix}
-2t \cos k& -2i\bar V\sin k \\
2i V \sin k & 2\Delta \cos k+\lambda
\end{pmatrix}}^{H (k)}
\begin{pmatrix}
c_{k\sigma}\\f_{k\sigma}\end{pmatrix}
\end{equation}
This model is represented schematically in Fig. \ref{fig:fig_TB}(a).

\subsection{Large $N$ limit} 

The mean-field treatment 
 replaces the 
hard constraint $n_{f}=1$ by an average $\langle  n_{f}\rangle =1$
at each site. This replacement becomes asymptotically exact in a large
$N$ extension of the model, in 
in which the fermions have  $N$ possible spin flavors,
$\sigma \in [1,N]$. Provided 
all terms in the Hamiltonian grow extensively with N, 
the path integral 
can 
be rewritten with an effective Planck constant $\hbar_{\hbox{eff}} =  1/N$ which
suppresses quantum fluctuations as $N\rightarrow \infty $ and  $\hbar_{\hbox{eff}}\rightarrow 0 $.
To scale the model so that the Hamiltonian grows extensively with $N$, we replace 
\begin{eqnarray}
&& 
J_H \rightarrow J_H/N,\quad\quad J_K \rightarrow J_K/N,
\\
&&
\sum_{j}\lambda (n_{f,j} -1)
\rightarrow 
\sum_{j}\lambda (n_{f,j} -Q).
\end{eqnarray}
where the last term imposes $n_{f,j} =Q$ rather than unity at each
site. We shall examine the case where $Q= N/2$, corresponding to a
particle-hole symmetric Kondo lattice.  
We shall restrict our attention to solutions 
where $\lambda=0$, which gives rise to an insulating state in which
both the conduction and f-bands are half-filled.

\subsection{Topological class D} The mean-field Hamiltonian (\ref{TB
model}) can be classified according to the periodic table of free
fermion topological phases\cite{Schnyder2008,Qi2008,Kitaev2009}. The
particle hole symmetry $\Xi:$ $\Xi H (k)\Xi^\dag =-H^T (k)$ 
is equivalent to the transformation 
$c_k\rightarrow c_{\pi -k}^\dag$, $f_{k}\rightarrow f_{\pi -k}\dg $. 
In the two band basis of
Hamiltonian (\ref{TB model}) $\Xi = \tau_z$, where
$\tau$ denotes a Pauli matrix acting in  orbital space. 
According to the periodic table, symmetric $\Xi$ corresponds to class D.

 One way to see the non-trivial topology is to
observe the evolution of the Hamiltonian throughout the Brillouin zone by
writing it as a vector in three dimensional space: $H (k) = \vec{h}(k)
\cdot \vec{\tau}+ \epsilon_0(k) $  with $\epsilon_0(k) =
(\Delta-t)\cos k$. For real $V$, 
\begin{equation}\nonumber
\vec{h}(k) = 
\left[
\begin{array}{c}
0 
\\2V\sin k\\
-(\Delta+t)\cos k 
\end{array} \right].
\end{equation}
At $k=0$ and $k=\pi$, the vector $\vec{h} (k)$ aligns along the $\hat
z$ axis: if the sign of the scalar product $\vec h(0) \cdot \vec
h(\pi)$ of the two vectors is positive, vector $\vec h (k)$ traces a
simply-connected path on the 2-sphere that may be contracted to a
point, so the phase is topologically trivial.  By contrast, a negative
sign corresponds to a topologically non-trivial
path that connects the poles of the sphere, 
indicating  the topological phase. This can be surmised as
\begin{equation}
(-1)^\nu  = \text{sign}(\vec h(0)\cdot \vec h(\pi)),
\end{equation}
where $\nu = 0$ for trivial and $\nu = 1$ for topological phases.  In our
model  $\vec h(0)\cdot \vec h(\pi) = -(t+\Delta)^2$ and hence $\nu = 1$
for any uniform solution with finite $V$. 

The consequence of the topological invariance can be seen in the non-zero
electric polarization $P$. A particle-hole transformation reverses the
polarization, and since the Hamiltonian is invariant under
this transformation it follows that $\Xi P\Xi^\dag = -P$, allowing
only two possible values of polarization: $P = 0$ or $P = e/2$ since
$P$ is defined modulo $e$.  This is in fact the topological index of
the chain.  P can be computed via the Berry connection $A_k = i
\langle u_{k}|\partial_k| u_{k}\rangle$\cite{kingsmithvanderbilt} of
the occupied bands, defined via periodic part of the Bloch function,
$u_{k}$.
\begin{equation} \label{polarization}
P =  {e}
\int_{0}^{2\pi} \frac{dk}{2\pi} 
 A_k = \left\{
  \begin{array}{l l}
    e/2 & \quad \text{topological}\\
    0 & \quad \text{trivial} \end{array} \right.  \end{equation} 
The validity of this relation depends on 
the use of 
a \textit{smooth} Berry connection
$A_k$, which usually requires that we carry out a 
a gauge transformation on the raw eigenstates. 
For example, consider the special case where
$t = \Delta = V$. The negative energy 
eigenstates then take the form $\psi_k =
(\cos(k/2),-i\sin (k/2))e^{i \phi (k)}$: choosing $\phi (k)=k/2$, 
the eigenstates then become continuous. 
If the orbital basis is centro-symmetric, 
the Berry phase only depends on the $\psi_k$: $A_k = i
\langle u_{k}|\partial_k| u_{k}\rangle =i \psi_{k}^\dag\partial_k
\psi_{k}$. Computing the Berry connection, we obtain
\begin{eqnarray}\nonumber
A_{k}& = ie^{-i k/2} \left( {\cos \frac k2}, {i\sin \frac k2}\right) 
 \partial_k 
\pmat{\cos (k/2)\cr -i \sin (k/2)}e^{i k/2} \cr
\nonumber
&= \frac{1}{2}(\cos^2 (k/2)+\sin^2 ( k/2)  ) = \frac{1}{2}
\end{eqnarray}
so that 
\begin{align}\nonumber
P = { e}\int_{0}^{2\pi} \frac{dk}{2\pi} 
 A_{k} = \frac{e}{2},
\end{align}
resulting in a non trivial half integer  charge (per spin component) 
on the edge.

\subsection{Edge states} 

The key property of a topological insulators
and superconductors is that at the particle-hole 
symmetric point, they develop zero energy edge states. 
A single non-degenerate state at
zero energy can not be shifted up (or down) because particle-hole
symmetry would then require 
at least two states with opposite energies, developing out of 
the single zero energy mode. 

\begin{figure}[h!]
\centering
(a) \ \includegraphics[height=0.15\linewidth]{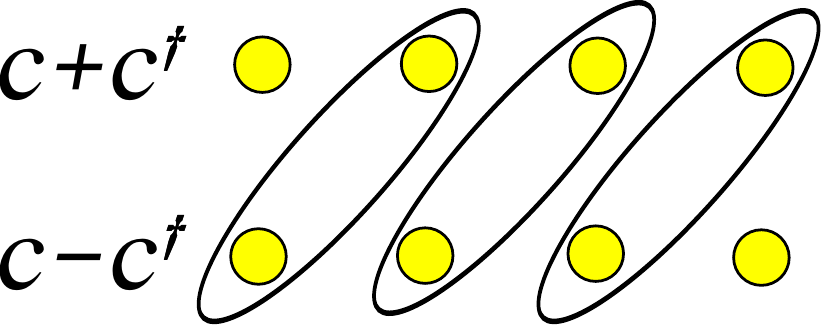}
 (b)\ \includegraphics[height=0.15\linewidth]{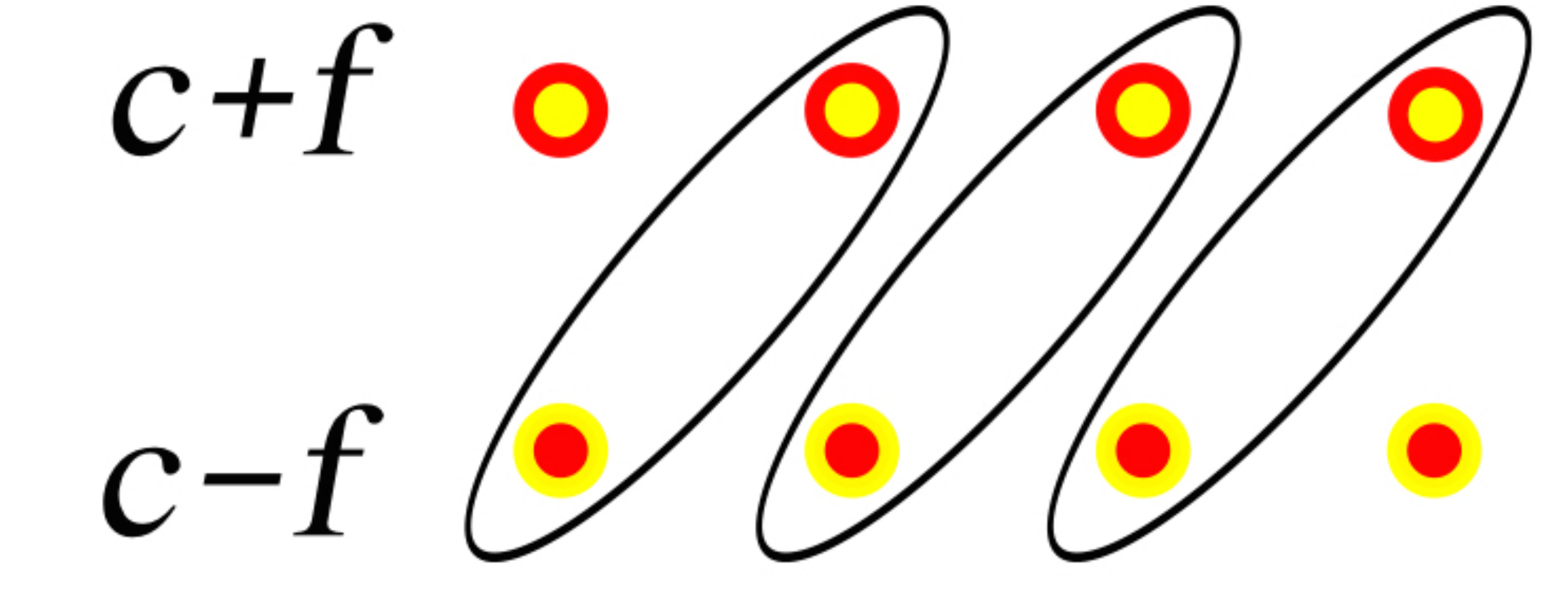}
\caption{(Color online) (a) Majorana decomposition of Kitaev model; (b) $t=\Delta = V$ limit of the tight binding model (\ref{TB model})}
\label{fig:fig_kitaev}
\end{figure}

Though our ultimate goal is to consider 
non-uniform mean field solutions, 
we begin by examining 
the form of topologically protected edge states for the mean-field
Kondo lattice (\ref{TB model}) with constant 
bond parameters. There is an interesting relationship with the 
topological Kitaev model\cite{Kitaev2001}, which we 
now bring out. The Kitaev model involves the formation of ``canted''
valence-bond solid between nearest neighbor Majorana fermions,
formed from symmetric and
antisymmetric combination of particle and holes form bonds, as shown in 
Fig. \ref{fig:fig_kitaev}a.  The edge states 
are then the Majorana fermions at the ends that are unable to form
bonds. We shall show that 
at the special point where the
all bond strengths are equal, the 
mean-field Kondo model involves the formation of  a similar canted
valence bond structure between antisymmetric and symmetric
combinations of 
of $f$ and conduction electrons 
as shown in Fig. \ref{fig:fig_kitaev}b.

To demonstrate the edge state wave function we can choose $f$ and $c$ hopping to be equal $t = \Delta$, keeping the hybridization a free parameter. 
The Hamiltonian (\ref{TB model}) can be rewritten in the following simple form:
\begin{eqnarray}\label{MF_kitev}
H_{(\Delta=t)} &=& (\Delta+V)\sum_{j,\sigma} (s\dg _{j+1\sigma}  a _{j\sigma} 
+
\mbox{H.c.})\cr
&+&
(\Delta-V)\sum_{j,\sigma} 
(s\dg_{j-1\sigma}a_{j\sigma} +{\rm H.c})
,
\end{eqnarray}
where
\begin{eqnarray}
a_{j\sigma} = (f_{j\sigma}-c_{j\sigma})/\sqrt2,\\
s_{j\sigma} =  (f_{j\sigma} + c_{j\sigma})/\sqrt2.
\end{eqnarray}
The two terms in Hamiltonian (18) correspond to  ``right facing'' and
``left facing'' bonds between a chain of ``a'' and ``s'' sites. 
In the particular limit that $\Delta =V$, the Hamiltonian consists
entirely of right-facing bonds, as illustrated in Fig. 3b, with edge
on the left and right  composed of symmetric and antisymmetric
combinations of conduction and ``f'' electrons. 
Note that at first glance this model breaks inversion symmetry, but in fact there is an additional $U(1)$ gauge invariance for $f$ electrons: the phase of $f$ can be rotated, effectively interchanging between antisymmetric $a_j$ and symmetric $s_j$ operators.

For all values of $V$ and $\Delta$, the zero-mode $\psi_0$   can be found solving  $H_{(\Delta=t)}\psi_0 = 0$ with the ansatz
\begin{equation}
   \psi_0 =\sum_{j} v_j s_j^\dag +u_j a^\dag_j  \\
\end{equation}  
 Solving for $\{u_j, v_j\}$ in
case  $(V \Delta) >0$ we can find left and right  edge solutions. The
left-hand edge state is given by
\begin{align}\label{kitaev_GS}
\begin{split}
& u_j=0
\\
& v_j =   \left\{
  \begin{array}{l l}
     \left(  {V +\Delta\over V- \Delta}\right)^{(j-1)/2} & \quad \text{odd}\\
    0 & \quad \text{even}
  \end{array} \right.
  \end{split}
\end{align}
Hence, unless hybridization $V$ or effective hopping $\Delta$ is zero
the decay is exponential. The fact that $v_{even} = 0$ is due to
particle hole transformation: one can show that the zero-mode of
bipartite lattice  is defined only on one sublattice in a one-dimensional finite chain.

\section{Mean field solution}
We now consider a finite slab of material, examining the departures 
in $V$ and $\Delta $ which develop in the vicinity of the boundaries a
phenomenon we refer to as ``Kondo band-bending''. 
The allowed values of $V_{j}$ and $\Delta_{j}$ are determined by the 
self-consistency equations 
\begin{align}\label{self_cons}
\begin{split}
& V_j = -J_K{(j)} \left \langle  (c_{j+1,\sigma}- c_{j-1,\sigma})^\dag f_{j,\sigma}\right \rangle,
\\
&\Delta_j =-  J_H\left \langle  f_{j,\sigma}^\dag f_{j+1,\sigma}\right \rangle,
\end{split}
\end{align}
These equations derive from the requirement that the acton is
stationary with respect to $V_{j}$ and $\Delta_{j}$ at each site.
The phase diagram is determined by 
the values of $J_{H}$, $J_{K}$ and the edge parameter $\alpha $. 
To explore the parameter space 
we carried out a series of 
a numerical calculations in which we seeded 
inhomogeneous order parameters $V_j$ and $\Delta_j$, iterating the
self-consistency conditions until a convergent solution was found. 

%
%It is not at all clear that uniform (in space) $V,t,\Delta$ can provide a solution to (\ref{decrete_MF}). In fact, nowhere on the phase diagram ($J_K,J_H$) it does so. We introduce another intuitive parameter that can change the above statement. We define $\alpha$ to characterize the boundary renormalization of Kondo coupling $J_K$.
%\begin{equation}\label{alpha}
%J_K^{boundary} =\alpha J_K^{bulk}  
%\end{equation}
%One of the interesting results of this work is how the properties of end-states change with variation of alpha. For instance, the ground state of $V = t = \Delta$ occurs for $\alpha = 2$. This scenario we refer to as ``Kitaev'' point as explained below.

\begin{figure}[h!]
\centering
\includegraphics[width=0.65\linewidth]{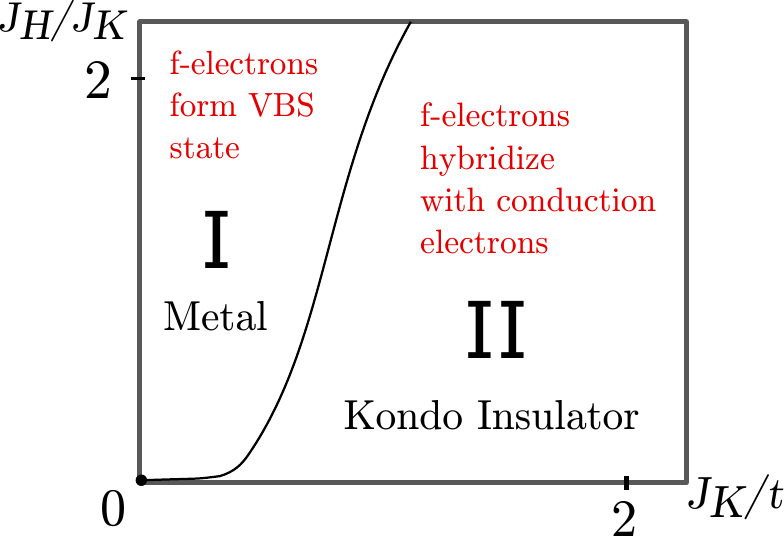}

\caption{Schematic phase diagram of the bulk ground state contains metallic phase (I) and insulating phase (II). Phase II can be further divided into surface phases depending on the properties of its surface states see Fig. \ref{fig:fig_phases 2}.}
\label{fig:fig_phases}
\end{figure}

To examine the bulk properties, we began by imposing 
periodic boundary conditions. Using this procedure, we identified 
two bulk phases: a Kondo insulator and a 
metallic valence bond solid. 
The results of mean field calculations with periodic boundary
conditions are presented in Fig. \ref{fig:fig_phases}. The bulk 
phase diagram is of course independent of the edge parameter $\alpha$.

We proceed with open boundary conditions and examine the nature of the bound states solutions that develop at the ends where mean field parameters depart from the bulk.

\subsection{Phase I: Metallic VBS state}

\begin{figure}[h!]
\begin{tabular}{c@{\hskip 12pt}c}
\includegraphics[width=0.45\linewidth]{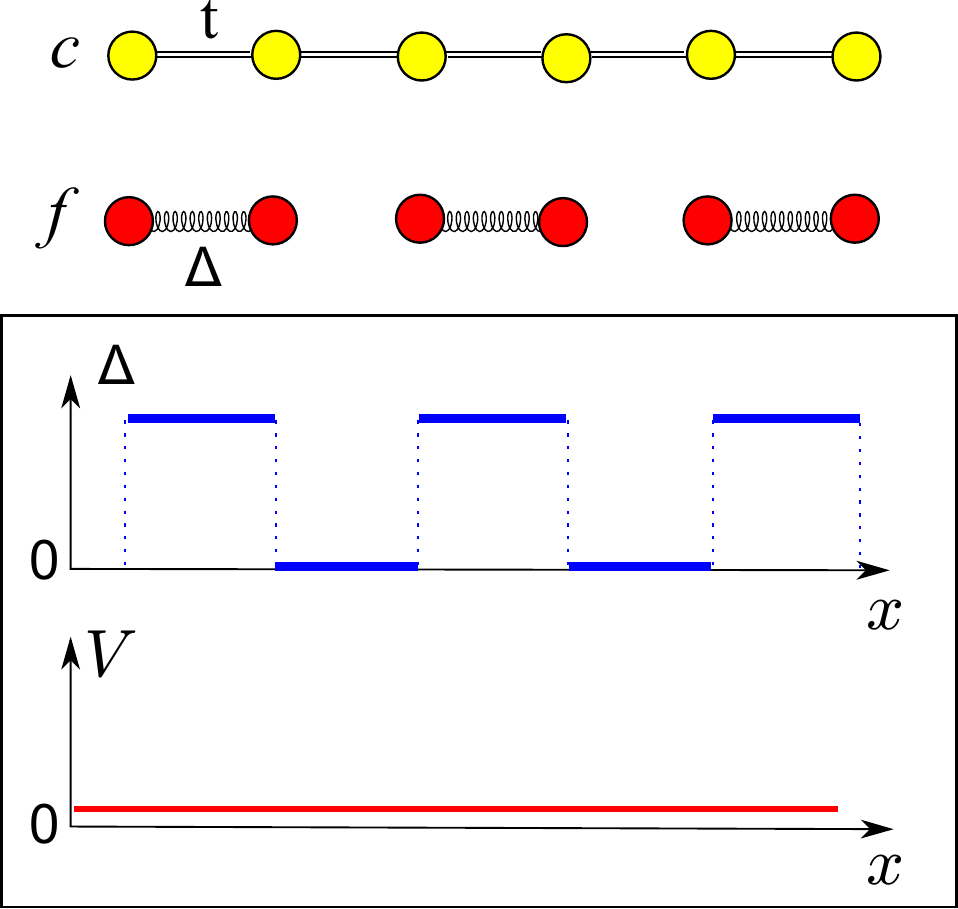}  &
\includegraphics[width=0.45\linewidth]{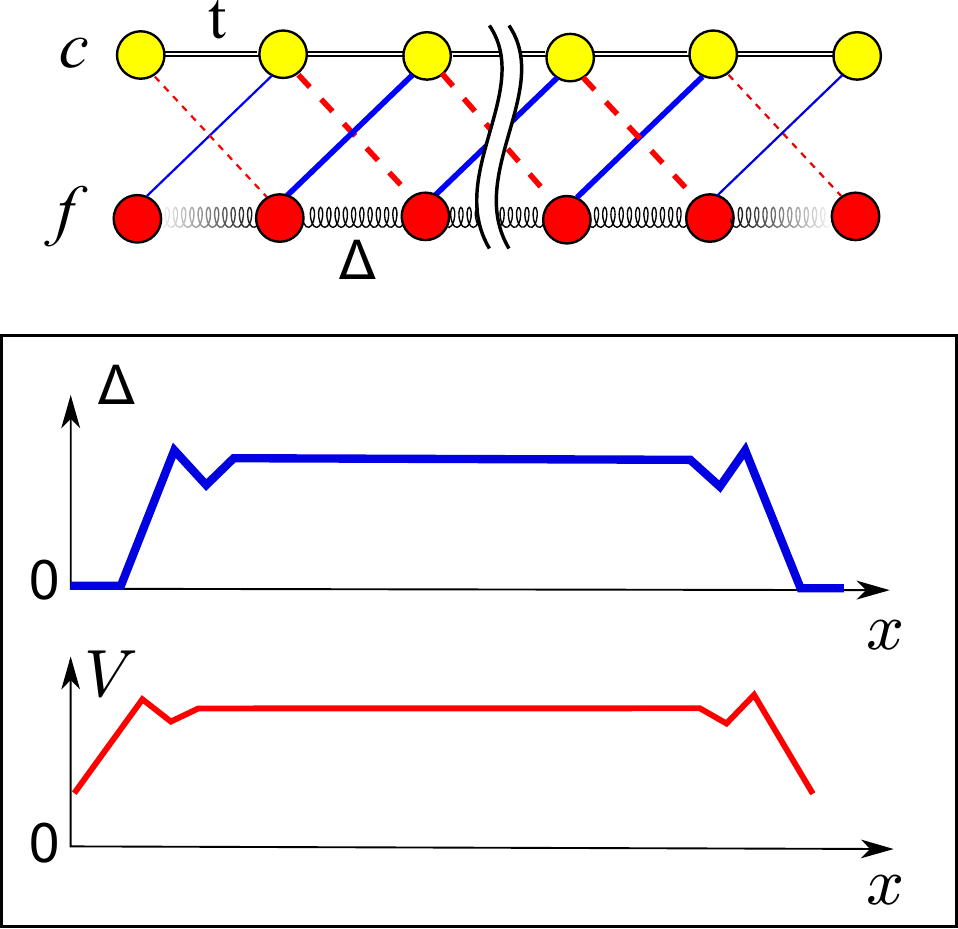}\quad
\\
(a) & (c) \\
\\
\includegraphics[width=0.25\linewidth]{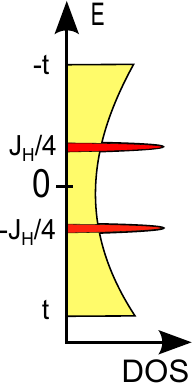}\hfill~& 
\includegraphics[width=0.45\linewidth]{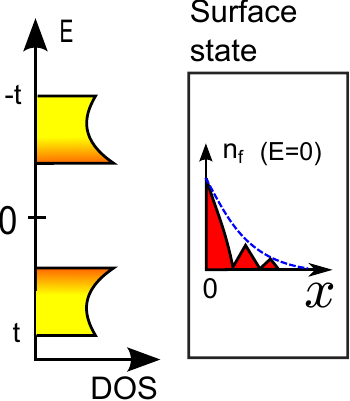}\\
\\
  (b)&  (d)
\end{tabular} 
\vspace{-.05cm}
	\caption{Cartoon representation of two distinct phases. Spatial dependence of $V_j$ and $\Delta_j$ and density of states (DOS) for phase I and II in Fig. \ref{fig:fig_phases}.Phase I (a and b): Metallic states with suppressed hybridization. It does not support surface states. Phase two (c and d): Kondo topological insulator that support surface states. Inset in d is the profile of a typical surface state decaying into the bulk as a function of distance. }
\label{fig:fig_dimer}
\end{figure}

In this phase the RVB order parameter $\Delta_j$ becomes
an alternating function of space, while $V_j$ is zero. Consequently,
the f-electrons form a valence bond solid (VBS) state, co-existing
with the unperturbed conduction sea. 
Dispersionless ``spinon'' bands above and below the Fermi
energy, 
as shown in see Fig.  \ref{fig:fig_dimer}b. The gap between f-states is provided
by the amplitude of $\Delta_j$ (justifying the notation) which is in
turn equal to $J_H/4$. 
The metallic VBS phase is summarized in
Fig. \ref{fig:fig_dimer}a,b.

The metallic phase does not have surface states and behaves the same way for open and closed boundary conditions.
We found  VBS state to be the lowest energy configuration in the left part of phase diagram as shown in Fig. \ref{fig:fig_phases}. 

Since there are two degenerate configurations of the VBS, 
one of the important classes of excitation of this state is a
domain-wall soliton formed at the interface of the two degenerate
vacua.  In an isolated VBS, such as the ground-state of the Majumdar
Ghosh model, or the Su-Schrieffer-Heeger model, such solitons are
spin-1/2 excitations. However, in the 1D Kondo lattice, the Kondo
interaction is expected to screen such isolated spins, forming a
p-wave Kondo singlet exciton. In the metallic VBS, these
solitonic excitons  will be gapped excitations.
However, as 
the Kondo coupling grows, at some point the 
excitons will condense, and at this point the VBS melts, forming 
a topological Kondo insulator. 

\subsection{Phase II: Kondo insulator}

In the Kondo insulating phase the hopping $\Delta_j$ and $V_j$ are
both finite in the bulk and generally suppressed at the ends of the
chain. This  gapped heavy Fermi liquid is stabilized by large Kondo
coupling as shown in Fig. \ref{fig:fig_dimer}(c,d). We find that that
the Kondo insulating phase exhibits two different kinds of boundary
behavior.  In the
mean field theory, we can characterize these two phases by the
fractional conduction electron character $n_{c}\in [0,1]$ of the edge
state. The first is adiabatically
connected to the ``Kitaev point'' (see below) in which conduction
electrons and composite f-electrons hybridize to form the 
a surface state with $n_{c}>0$. 
In the second state, the edge state is a purely
localized spin, unhybridized with the conduction electrons ($n_{c}=0$).

\subsubsection{``Kitaev'' point}

For general values of $\{J_K,J_H,\alpha\}$ there is no 
 analytic solution. However at the point where 
$\{J_K,J_H,\alpha\}= \{2,4,2\}$ 
$\Delta_j = V_j=t$ are constants in space. We now show that at this point
each spin component of the mean-field theory corresponds to a 
two copies of the Kitaev chain, with a single fermionic zero mode at
each boundary, 
as can be seen
from the form of the wave function in equation (\ref{kitaev_GS}). We
refer to this particular point in the phase diagram as 
the  'Kitaev point'.

At this point, following (\ref{MF_kitev}), the mean-field Hamiltonian
takes the form 
\begin{equation}\label{kitpoint}
H_{(\Delta=t)} =2t\sum_{j=1}^{L-1 }\sum_{\sigma} (s\dg _{j+1\sigma}  a _{j\sigma} 
+\mbox{H.c.}),
\end{equation}
where $s_{j\sigma }$ and $a_{j\sigma }$ are the symmetric and
antisymmetric combination of states $\frac{1}{\sqrt{2}} (f_{j\sigma
}\pm c_{j\sigma })$. 
This Hamiltonian commutes with the zero modes 
\begin{equation}\label{}
[H,s_{1\sigma }]=0, \qquad [H,a_{L\sigma }]=0
\end{equation}
so for each spin, there is one fermionic zero mode per edge, each
involving a hybridized combination of conduction and f-electrons with
$n_{c}=\frac{1}{2}$. 
To see the connection with the Kitaev model, 
we divide both $s_{j\sigma }$ and $a_{j\sigma}$ into two
Majorana fermions as follows, 
  \begin{eqnarray}
  s_{j\sigma } = 
\frac{1}{\sqrt{2}} (\gamma^{1}_{j\sigma }+ i
  \gamma^{2}_{j\sigma }), \qquad 
  a_{j\sigma } = \frac{1}{\sqrt{2}} (\gamma^{3}_{j\sigma }- i
  \gamma^{4}_{j\sigma }).
  \end{eqnarray}
  Using (\ref{kitpoint}), the Hamiltonian now splits into two independent
components, 
  \begin{equation}
  H_{(\Delta =t) } = -i2t \sum_{j=1 }^{L-1}  \sum_{\sigma }  
(\gamma^1_{j+1} \gamma^4_{j} + \gamma^2_{j+1}\gamma^3_{j}),
  \end{equation}
corresponding to a pair of Kitaev chains per spin component. This is
natural, because each Kitaev chain has one Majorana zero mode per
edge. Since a pair of Majoranas make one normal fermion, this
corresponds to one fermionic  zero mode per edge. 

\begin{figure}[h!]
\includegraphics[width=0.45\linewidth]{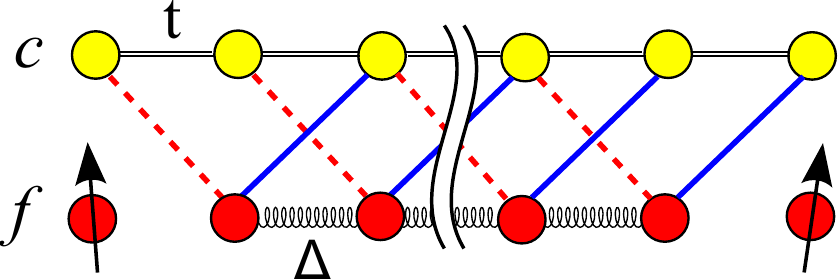} 
\caption{(Color online) Magnetic phase (red color in Fig \ref{fig:fig_phases 2}). }
\label{fig:fig_magnetic}
\end{figure}

\subsection{Magnetic Edge state. }\label{}

In the magnetic edge state, the boundary spins do not undergo the
Kondo effect, forming an unhybridized magnetic edge state.  If the
boundary parameter $\alpha =1$, the Kondo temperature at the boundary
is smaller than in the bulk, because the terminal boundary spins have
only one nearest neighbor.  This means on cooling, that the boundary
Kondo interaction is unable to scale to strong coupling before a gap
develops in the bulk, leading to an unquenched boundary spin.  When
$\alpha >1$, the Kondo effect is able to develop at the boundary,
occurs at the boundary, provided $J_{K}$ is not at weak coupling.
At smaller values of $J_{K}$, the decoupled magnetic
phase develops, denoted by the red region in Fig. 7 (a). 
In this phase, there is no hybridization of the edge
state with the bulk conduction electrons ($n_{c}=0$), and the
topological edge state disappears.

\begin{figure}[h!]

%\begin{picture}(0,0)
%\put(0,140){\large (a)}
%\put(0,-10){\large (b)}
%\put(120,-10){\large (c)}
%\end{picture}
\includegraphics[width=0.95\linewidth]{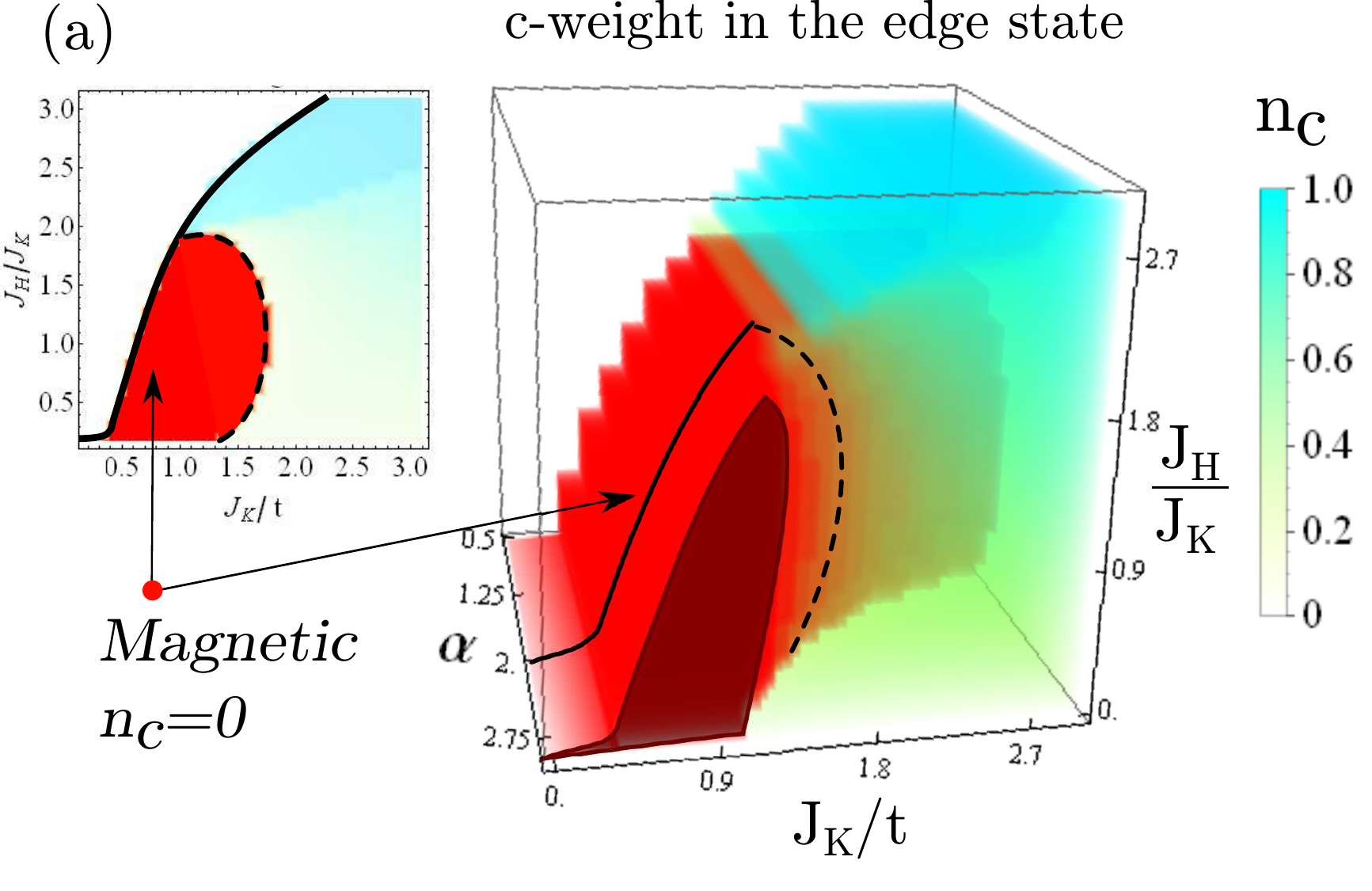}
\includegraphics[width=0.99\linewidth]{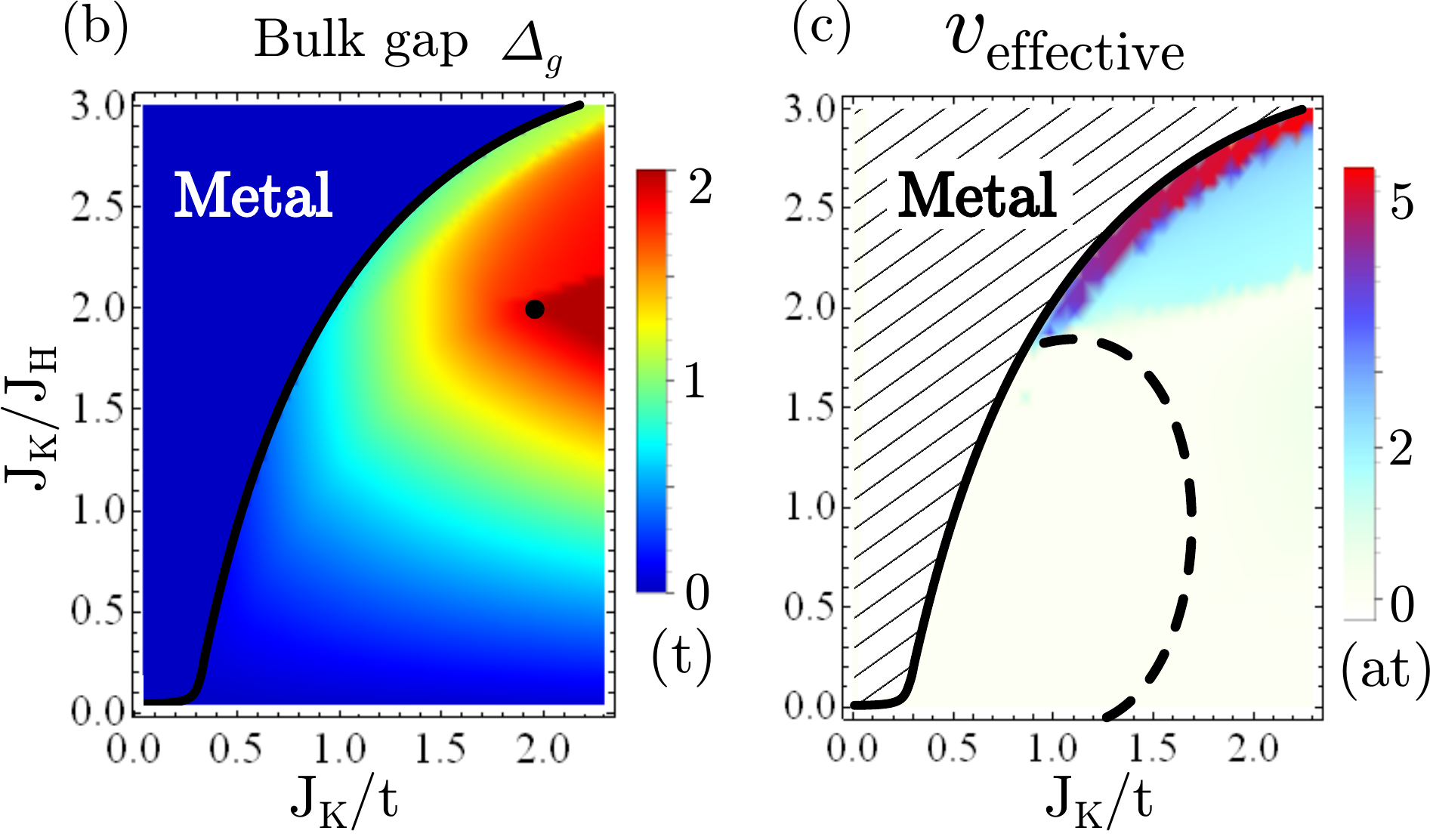}
\caption{(Color online) The surface phase diagram. With ans addition
of renormalized Kondo coupling at the surface $\alpha =
J_K^{boundary}/ J_K $. (a) Red color represent pure f-states with no
c-electron mixing, $n_c = 0$. Inset is the cut along $\alpha =2$. (b)
(b) is the bulk gap $\Delta_g$ in the units of $t$ for $\alpha =2$,
identifying the the ``Kitaev point'' as a black dot.  (c) displays the
inferred group velocity measured in units of $a t$, derived from the
measured penetration depth of the edge states using $v= \Delta_g\xi$.
The mean field calculations were done on a chain of 70 unit cells. 
} \label{fig:fig_phases 2}
\end{figure}

\section{Results and Discussion}

\subsection{Where are the light surface states? }\label{}

One of the interesting features of current experiments on the Kondo
insulator SmB$_{6}$, is that the putative topological surface states
seem to involve high-velocity quasiparticles, rather than the heavy,
low-velocity particles predicted by current theories.
Our mean-field results on the one-dimensional p-wave Kondo chain
suggest that this may be because the change in character of the Kondo
effect at the boundary leads to edge states with a 
large conduction electron component. 

For a relatively high magnetic interaction, $J_H$, the one-dimensional
edge states in our mean-field treatment 
develop majority  conduction electron character,
forming ``light'' edge states which penetrate deeply into the bulk.

In a non-interacting topological insulator, the transition to 
to a topologically trivial phase occurs via a quantum phase transition
in which the bulk gap closes. In this case, the penetration 
penetration depth grows with inverse proportion to the bulk gap
$\Delta_g$.
\begin{equation}\label{corr_length}
\xi =v_F/\Delta_g
\end{equation}
However, in the p-wave Kondo chain, 
the transition to a metallic VBS is a first order transition at which the 
the bulk gap remains finite. In this case, 
the rapid growth in the penetration depth of the edge state 
is associated with an increase in the conduction character, driving 
an enhanced  group-velocity of the edge-states
(\ref{corr_length}).  
This is a novel and interesting
consequence of the response of the Kondo effect to the boundary - ``Kondo band-bending''. 

To demonstrate this behavior, we have carefully examined the
properties of the edge states in our model.
The phase diagram showing the evolution in the conduction
character of the end states character is shown in
Fig. \ref{fig:fig_phases 2}a.
Within the bulk topological insulator phase, 
the character of the edge states varies
dramatically, ranging from equal f- and c-
character at the to 
to edge states of predominantly conduction electron
character near the first order boundary. 
Fig. \ref{fig:fig_phases 2}b shows the dependence of the insulating
gap $\Delta_g$, showing that it remains finite at the first order
phase boundary to the VBS metal. 

We can estimate the 
the effective velocity
of the edge states by combining the measured coherence length of the
edge state and the bulk gap, according to 
\begin{equation}\label{}
v_{\hbox{effective}}= \Delta_{g}\xi
\end{equation}
This quantity is found to increase dramatically near the first order
boundary into the metallic VB state (see \ref{fig:fig_phases 2}c),
unlike a non-interacting topological insulator, here the increase in
$\xi$ is due to a rapidly increasing amount of conduction character in
the edge-states, and is not accompanied by a gap closure, so that the
effective velocity of the edge states $v_{\hbox{effective}}$ rises considerably.

\subsection{Strong-coupling Ground state wave function} 

An alternative way to understand 
a Kondo insulator is 
through the character of its strong-coupling wavefunction. 
In an conventional  Kondo insulator, the strong coupling 
ground-state is an array of Kondo singlets. 
If we write
\begin{equation}\label{}
A\dg_{j} = \sum_{\sigma =\pm \frac{1}{2}} f\dg_{j\sigma }c\dg_{j,
-\sigma}{\rm sign } (\sigma)
\end{equation}
then the strong coupling ground-state of the s-wave Kondo insulator is
simply a  valence bond solid of Kondo singlets:
\bxwidth=1.8truein
\begin{eqnarray}\label{l}
\vert  KI\rangle  &=& \prod_{j =1}^{L}A\dg_{j}\vert 0\rangle.
\cr
&=& \raisebox{-0.2truein}[0cm][0cm]{\frm{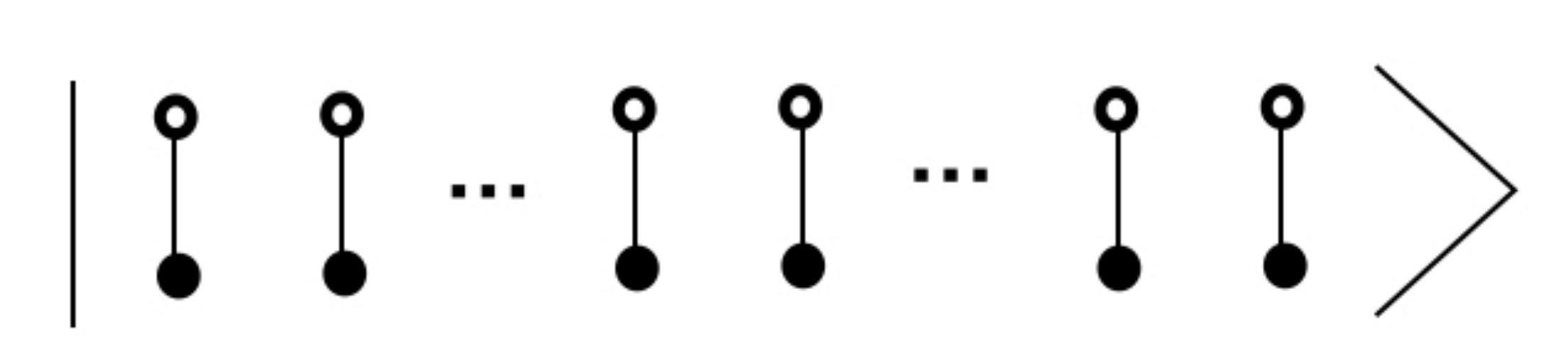}}
\\ \nonumber
\end{eqnarray}
where a line denotes a valence bond between a conduction electron
(open circle) and a local moment (closed circle).

What then is the corresponding ground-state for the topological Kondo insulator? 
We can construct variational wavefunctions for the topological Kondo
insulator by applying a Gutzwiller projection to the mean-field
ground-state. Unlike the s-wave Kondo chain, to preserve the
topological ground-state, we need to consider large values for both
the Kondo and the Heisenberg coupling. An interesting point to
consider is the Kitaev point, where the singlet structure of the
mean-field ground-state becomes highly local. By projecting the
mean-field ground-state we obtain
\begin{equation}\label{}
\vert TKI\rangle  = P_{G} \prod_{j} Z_{j}\vert 0\rangle 
\end{equation}
where 
\begin{equation}\label{}
Z_{j}= \sum_{\sigma = \pm \frac{1}{2}} a\dg_{j\sigma }s\dg_{j+1, -
\sigma }{\rm sign \sigma }
\end{equation}
with $a\dg_{j\sigma }= (f\dg_{j\sigma }+ c\dg_{j\sigma })/sqrt{2}$
and $s\dg_{j\sigma }= (f\dg_{j\sigma }- c\dg_{j\sigma })/sqrt{2}$
as before, whereas $P_{G}= \prod_{j} (n_{f\uparrow} (j)-n_{f\downarrow
} (j))^{2}$. Now the valence bond-creation operator
\begin{eqnarray}\label{l}
Z\dg _{j} &=& \frac{1}{2} \sum_{\sigma }
(f\dg_{j\sigma }f\dg_{j- \sigma } + f\dg_{j\sigma }c\dg_{j+1,-\sigma
}\cr
&-& c\dg_{j\sigma }f\dg_{j+1,-\sigma }
- c\dg_{j\sigma }c\dg_{j+1,-\sigma }){\rm sign} (\sigma )
\end{eqnarray}
is non-local. The projected wavefunction of the TKI now involves a
multitude of configurations forming a one-dimensional resonating
valence bond (RVB) state between the local moments and conduction
electrons.  Schematically, 
\begin{eqnarray}\label{l}
\vert TKI\rangle =
 \sum
 \raisebox{-0.2truein}[0cm][0cm]{\frm{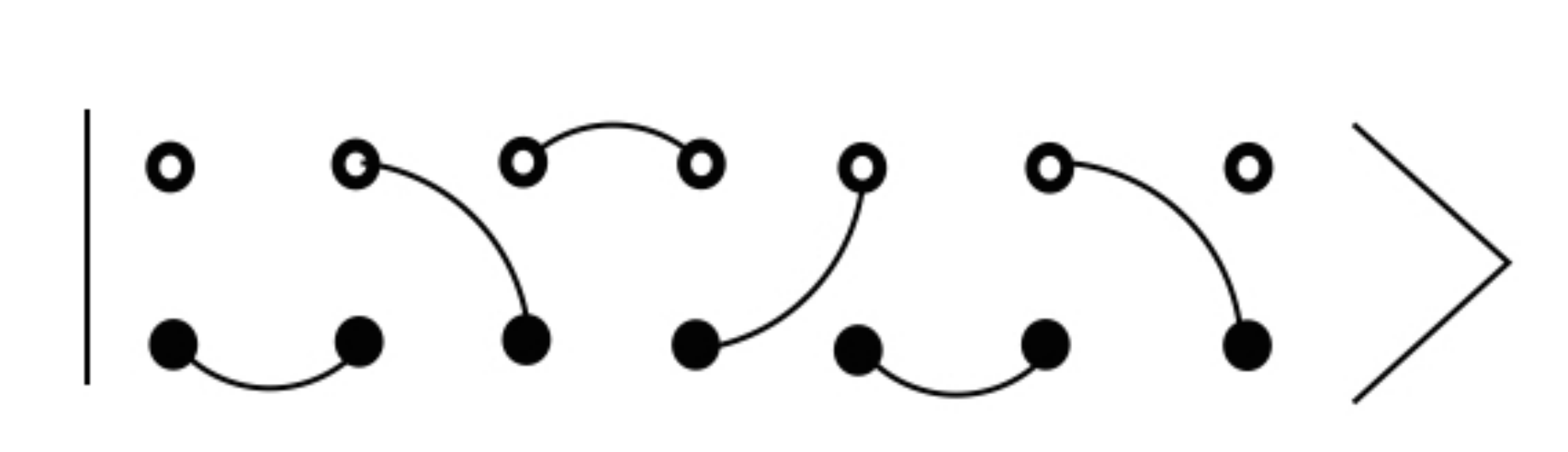}},
\\  \nonumber
\end{eqnarray}
where we associate a minus sign with left-facing Kondo singlets and
conduction electron pairs. 
In this picture, the edge-states correspond to unpaired spins or
conduction electrons at the boundary. 
\begin{eqnarray}\label{l}
\vert \hbox{edge},\sigma \rangle &=&
P_{G}s\dg_{1\sigma }\prod_{j}Z\dg _{j}\vert  0 \rangle  \cr
&=& \sum
 \raisebox{-0.2truein}[0cm][0cm]{\frm{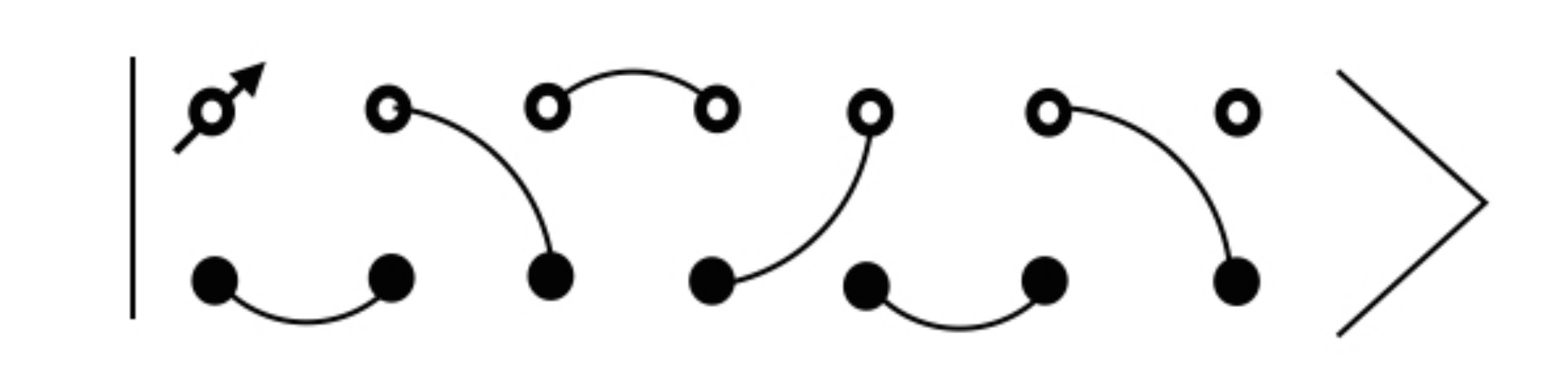}}\\
\nonumber
\cr
&+& \sum\ 
 \raisebox{-0.2truein}[0cm][0cm]{\frm{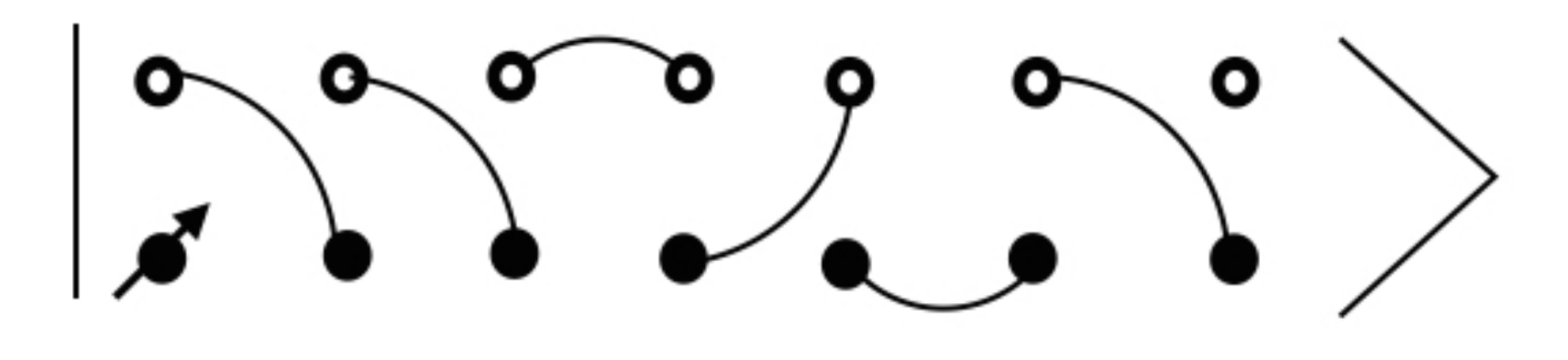}}
\\
\nonumber
\end{eqnarray}
At the current time, except in the large $N$ limit, 
we do not yet know if there is a particular combination of
$J_{K}$, $J_{H}$ and hopping $t$ for which the short-range RVB
wavefunction is an exact ground-state for the TKI. 

\subsection{Further outlook}

%%  What are the points here, aside from recap? 
%%
%%  Absorb the slow/fast into earlier section? 
%%
%%  Future work? 
%%
One of the interesting unsolved questions of is why different methods
of growing SmB$_6$ sometimes suppress the topological surface states.
On the one hand, when grown in Al flux, SmB$_{6}$ has robust surface
states with a low temperature plateau conductivity, whereas the
crystals produced with the floating zone method exhibit no plateau
conductivity, even though the samples are thought to be cleaner
\cite{Phelan2014}. Based
on our simple one-dimensional model, we speculate that this may
because the ordered surface supports localized magnetic moments which
in three dimensions, magnetically order.  By contrast, for reasons not
currently clear, the Al flux grown samples appear to 
sustain non-magnetic surface states, possibly due to a
valence shift at the surface, giving rise to topological surface
states. 
A more detailed understanding of the situation awaits an extension of
our current results to a three dimensional model along the lines of \cite{Yakovenko2012}.   This is work that
is currently underway. 

Finally, we note that the model we have discussed in this paper 
can also be engineered in a framework of ultracold atoms where a 
double well lattice potential is populated with mobile atoms in 
s and p orbitals\cite{Vincent2013}. This may provide a setting for a direct
examination of the 1D edge states.

\section{Acknowledgments}
The authors gratefully acknowledge discussions with
Onur Erten, Karen Hallberg and Tzen Ong. 
This work was supported by DOE Grant No. DE-FG02-99ER45790.

\end{document}